\documentclass[11pt,prd,letterpaper,tightenlines,superscriptaddress,showpacs]{revtex4}
\usepackage{amsmath,amssymb,amsfonts,graphicx,dcolumn}
\usepackage{epstopdf}
\usepackage{color}
\usepackage[utf8]{inputenc}
\usepackage{amsthm}
\usepackage{fontenc}
\usepackage{slashed}
\usepackage[dvips]{hyperref}
\usepackage{graphics}
\RequirePackage{slashed}
\usepackage{cancel}
\usepackage[normalem]{ulem}

\usepackage{array,booktabs}
\newcolumntype{C}{>{$\displaystyle}c<{$}}
\begin{document}

\pacs{}
\keywords{}

\title{Double parton scattering in $pA$ collisions
at the LHC revisited}

\author{Boris Blok}
\email{blok@physics.technion.ac.il}
\affiliation{ Department of Physics, Technion, Israel Institute of Technology, Haifa, 32000 Israel}
\author{Federico Alberto Ceccopieri}
\email{federico.ceccopieri@hotmail.it}	
\affiliation{ Department of Physics, Technion, Israel Institute of Technology, Haifa, 32000 Israel}
\affiliation{IFPA, Université de Li\`ege, B4000, Li\`ege, Belgium}

\begin{abstract}
\vspace{0.5cm}
We consider the production of $W$-boson plus dijet, $W$-boson plus b-jets and same sign $WW$ via double parton scattering (DPS)   in $pA$ collisions at the LHC and evaluate the corresponding cross sections. 
The impact of a novel DPS contribution pertinent to $pA$ collisions is quantified. 
Exploiting the experimental capability of performing measurements differential in the impact parameter
in $pA$ collisions, we discuss a method to single out this novel DPS contribution. 
The method  allows the subtraction of the  leading twist (LT)  single parton scattering background and it gives access in a very clean way to double parton distribution functions in the proton. 
We calculate leading twist  and DPS  cross sections and study  
the dependence of the observables on the cuts in the jets phase space.
In the $Wjj$ channel  the observation 
of DPS is possible with data already accumulated in $pA$ runs and the situation will greatly improve for the next high luminosity runs. For the $Wb\bar{b}$ final state, the statistics within the current data is too low,
but there is possibility to observe DPS in this channel in future runs, albeit with much reduced sensitivity than in  $Wjj$  final state. 
 Finally the DPS observation in the same sign $WW$ channel will require either significant increase of integrated luminosity beyond that foreseen in next $pA$ runs or improved methods for $W$ reconstruction, along with its charge, in hadronic decay channels. 
\end{abstract}

\maketitle

\section{Introduction}
\label{Sec:Intro}
\par The flux of incoming partons in hadron-induced reactions increases with the collision energy so that multiple parton interactions (MPI) take place,
both in $pp$ and $pA$ collisions. The study of MPIs started in eighties in Tevatron era~\cite{TreleaniPaver82,mufti}, both experimentally  and theoretically.
Recently a significant progress was achieved in the study of MPI, in particular of double parton scattering (DPS). From the theoretical point  of view a new self consistent 
pQCD based formalism was developed both for 
$pp$~\cite{stirling,BDFS1,Diehl,stirling1,BDFS2,Diehl2,BDFS3,BDFS4,Diehl:2017kgu,Manohar:2012jr} and $pA$ DPS collisions~\cite{BSW} (see~\cite{book} for recent reviews). From the experimental point of view, among many DPS measurements performed recently, the one in the $W$+dijet final state
is of particular relevance for the present analysis. The corresponding cross section was measured in $pp$ collisions both by ATLAS and CMS~\cite{ATLAS_WJJ,CMS_WJJ}. Moreover  recent observations of double open charm ~\cite{Belyaev,LHCb,LHCb1,LHCb2} and 
same sign $WW$ ($ssWW$) production~\cite{Sirunyan:2019zox} clearly show the   existence of DPS interactions in $pp$ collisions. 
\par  The MPI interactions play a major role in the Underlying Event (UE) and thus are taken into account in all MC generators developed for the LHC~\cite{pythia,herwig}.
On the other hand the study of DPS will lead to understanding of two parton correlations  in the nucleon. In particular 
the DPS cross sections involve new non-perturbative two-body  quantities, the so-called two particle Generalised  Parton Distribution Functions ($_2$GPDs), which encode novel  features of the non-perturbative nucleon structure.
Such distributions have the potential to unveil two-parton correlations in the nucleon structure~\cite{calucci,Rinaldi:2018slz} and to give access to information complementary to the one obtained from nucleon one-body distributions. 

\par The study of MPI and in particular of the DPS reactions in  $pA$ collisions is important for our understanding of  MPI in $pp$ collisions and it constitutes a  benchmark of the theoretical formalism available for these processes. On the other hand the MPI in $pA$ collisions may play an important role in underlying event (UE) and high multiplicity events in $pA$ collisions. Moreover it was argued in Ref.~\cite{BSW} that they are directly related to longitudinal parton correlations in the nucleon.
\par The theory of MPI and in particular DPS in $pA$ collisions was first developed in~\cite{Strikman:2001gz}, where it was shown that there are two DPS contributions at work in such a case. 
\begin{figure}[t]~
\includegraphics[scale=0.8]{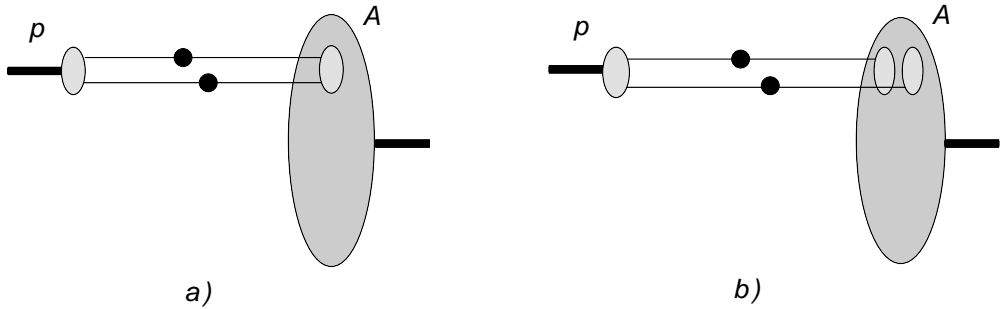}
\caption{\textsl{Pictorial representation of DPS process in $pA$ collisions via a) DPS1 and b) DPS2 mechanisms. The light grey blobs indicate nucleons, darker grey ones the nucleus and black ones the hard interactions.}}
\label{plot:DPS_pictorial}
\end{figure}
First, there is the so-called DPS1 contribution, depicted in the left panel of Fig.~(\ref{plot:DPS_pictorial}), in which two partons from the incoming nucleon interact with two partons in the target nucleon in the nucleus, making 
such a process formally identical to DPS in the $pp$ collisions. Next there is a new type of contribution, 
depicted in the right panel of Fig.~(\ref{plot:DPS_pictorial})
and often called DPS2, in which 
two partons from the incoming nucleon interact with two partons  each of them belonging to the distinct nucleons in the target nucleus located at the same impact parameter. Such a contribution is parametrically enhanced by a factor $A^{1/3}$ over the DPS1 contribution, $A$ being the atomic number of the nucleus.
\par The basic challenge in observing and making precision studies of DPS both in $pp$ and $pA$ collisions is the tackling the large leading twist (LT), single parton scattering (SPS) background. 
This problem is especially acute in $pA$ collisions where, due to several orders of magnitude lower luminosity relative to $pp$ collisions, rare DPS cross sections will suffer serious deficit in statistics~\cite{Helenius:2019uge,sde}. 
\par Recently  a new method was suggested~\cite{Alvioli:2019kcy} which could allow the observation of DPS2 in $pA$ collisions. It was pointed out that the DPS2 has a different dependence on impact parameter than SPS and DPS1 contributions.
Namely while the latter contributions are  proportional to  the nuclear thickness function $T(B)$, $B$ being the $pA$ impact parameter, the  DPS2 contribution is proportional to the square of $T(B)$. Therefore the cross section producing a given final state can be schematically written as:
\begin{equation}
\frac{d^2\sigma_{pA}}{d^2B}= 
\Big(\sigma^{LT}_{pA}+\sigma^{DPS1}_{pA}\Big) \frac{T(B)}{A} +\sigma_{pA}^{DPS2} \frac{T^2(B)}{\int d^2 B \, T^2(B)}\,,
\label{e1}
\end{equation}
where $T(B)$ is normalized to the atomic number $A$ of the nucleus. 
This observation gives the possibility to distinguish the DPS2
contribution in $pA$ collisions from both the LT SPS and DPS1 contributions that are instead linear in $T(B)$.
This approach was used in Ref.~\cite{Alvioli:2019kcy} to study two-dijets processes in $pA$ collisions. 
\par The purpose of the present paper is to investigate whether the latter approach can be used to observe the DPS2 process in $pA$ collisions  for the following  final states, ordered by decreasing cross sections: 
\begin{eqnarray}
pA &\rightarrow& W^\pm + dijets +X\,, \nonumber\\
pA &\rightarrow& W^\pm + b \bar{b}-jets +X\,, \nonumber\\
pA &\rightarrow& W^\pm + W^\pm +X\,. \nonumber
\end{eqnarray}
In all  considered channels one electroweak boson ($W^\pm$) is 
produced in one of the scatterings, which then leptonically decays into muon and a neutrino. 
A second scattering  in the same $pA$ collision produces the remaining part of the final state ($jj,b\bar{b},W^\pm$). The first process, as it emerges from our simulations, has the advantage of higher statistics which could allow 
the characterization of the DPS cross section. 
The second one has been discussed in detail in Ref.~\cite{Berger:2011ep} in $pp$ collisions and, despite the lower rate, its study is relevant since DPS contribution is an important background 
to new physics searches with the same final state. The third one is a gold channel DPS reaction but suffer from very low cross sections~\cite{{Kulesza:1999zh,Maina:2009sj,Gaunt:2010pi}}. 
\par In this paper we present the numerical estimates both 
for the  DPS cross sections  and  for the background LT contributions, the impact of the latter driving  the experimental capability 
to discover the DPS2 mechanisms in the current and in the future $pA$ runs at the LHC.
\par We show  that in the $Wjj$   final state  there is rather large number of events that allows to determine DPS2 already from data already recorded in $pA$ runs in 2016 at the LHC.
The situation will improve even more for the next runs for $pA$ runs at LHC scheduled for 2024.   For the $Wb\bar{b}$ final state  the statistics is too low to determine DPS2 in the current run,
but this may be possible in the future high luminosity runs, albeit with much reduced sensitivity than in  $Wjj$  final state.
On the other hand, $ssWW$ process suffers from a rather low statistics, even for the next runs. Nevertheless we expect we shall 
be able to observe it in the future runs if $W$ reconstruction techniques
will allow to establish the $W$ charge from its hadronic decays.
\par 
The paper is organised as follows. In Section \ref{Sec:theory} we review the theoretical framework on which are based our calculations. In the following three Sections
we present our results for each considered final state, $Wjj$,  $Wb\bar{b}$ and $ssWW$, respectively, and discuss the corresponding results. Our findings are summarised in the conclusion.

\section{Theoretical Framework}
\label{Sec:theory}
\noindent
The cross section for the production of final states $C$ and $D$ 
in $pA$ collisions via double parton scattering can be written as the convolution of the double $_2$GPDs of the proton and the nucleus, $G_p$ and $G_A$, respectively~\cite{BSW,Strikman:2001gz}:
\begin{equation}
\frac{d \sigma^{CD}_{DPS} }{d\Omega_C d\Omega_D}=\int \frac{d^2\vec \Delta}{(2\pi)^2}\frac{d \hat \sigma_{ik}^C(x_1, x_3)}{d\Omega_C}\frac{d \hat \sigma_{jl}^{D}(x_2,x_4)}{d\Omega_D}G_p^{ij}(x_1,x_2, \vec \Delta )G_A^{kl}(x_3,x_4, -\vec \Delta)\,.
\label{Eq:DPS_momentum_space}
\end{equation}
Two parton GPDs depend on the transverse momentum imbalance momentum $\vec \Delta$. The structure  and relative weight of different contributions to the nuclei $_2GPD$ was 
studied in detail in Ref.~\cite{BSW}, where it was shown that only two contributions survive:  the one that corresponds  to DPS1 mechanism and an other corresponding to DPS2.
\par Since our analysis will especially deal with impact parameter $B$ dependence of the cross section,  we find natural to rewrite 
Eq.~(\ref{Eq:DPS_momentum_space}) in coordinate space, introducing the double distributions $D_{p,A}$ which are the Fourier
conjugated of $G_{p,A}$ with respect to $\vec \Delta$. 
In such a representation these distributions admit a probabilistic interpretation and represent the number density of parton pairs with longitudinal fractional momenta $x_1,x_2$, at a relative 
transverse distance ${\vec{b}_\perp}$,  the latter  being the Fourier conjugated to $\vec \Delta$.
\par In the impulse approximation for the nuclei, neglecting possible corrections to factorisation due to the shadowing for large nuclei, and taking into account that $R_A \gg R_p$  for heavy nuclei, we can rewrite the cross section as~\cite{BSW,Strikman:2001gz}
\begin{multline}
\frac{d \sigma^{CD}_{DPS}}{d\Omega_1 d\Omega_2} = \frac{m}{2} 
\sum_{i,j,k,l} 
\sum_{N=p,n} \int d \vec{b}_\perp
\int d^2 B \, 
D^{ij}_{p}(x_1,x_2;\vec{b}_\perp) 
D^{kl}_{N}(x_3,x_4;\vec{b}_\perp)  T_N(B) 
\frac{d \hat \sigma_{ik}^C}{d\Omega_C} \frac{d \hat \sigma_{jl}^D}{d\Omega_D}\,,\\
+ \frac{m}{2} \sum_{i,j,k,l} \sum_{N_3, N_4=p,n}
\int d \vec{b}_\perp
D^{ij}_{p}(x_1,x_2;\vec{b}_\perp) 
\int d^2 B \,
f^{k}_{N_3}(x_3)
f^{l}_{N_4}(x_4)  T_{N_3}(B) T_{N_4}(B) 
\frac{d \hat \sigma_{ik}^C}{d\Omega_C} \frac{d \hat \sigma_{jl}^D}{d\Omega_D}\,.
\label{uno}
\end{multline}
Here $m=1$
if $C$ and $D$ are identical final states and $m=2$ otherwise,
$i,j,k,l = \{q, \bar q, g \}$ are the parton species contributing to the final states $C(D)$.
In Eq. (\ref{uno}) and in the following,
$d \hat{\sigma}$ indicates the partonic cross section for producing the final state $C(D)$, differential in the relevant set of variables, $\Omega_C$ and $\Omega_D$, respectively. 
The functions $f^i$  appearing in  Eq.~(\ref{uno}) 
are single parton densities and the subscript $N$ indicates nuclear  parton distributions.  The double parton distribution $D_N$ is the double GPD for the nucleon  bound in the nuclei,
once again calculated in the mean field approximation.

\par Partonic cross sections and parton densities do additionally depend on factorization and renormalization scales whose values are set to appropriate combination of the large scales occuring in final state $C$ and $D$. 
\par The nuclear thickness function $T_{p,n}(B)$, mentioned in the Introduction and appearing in Eq.~(\ref{uno}), is obtained integrating the proton and neutron densities $\rho_0^{(p,n)}$ in the nucleus over the longitudinal component $z$
\begin{equation}
\label{thickness}
T_{p,n}(B)=\int dz \rho^{(p,n)}(B,z)\,,
\end{equation}
where we have defined $r$, the distance of a given nucleon from nucleus center, in terms of the impact parameter $B$ between the colliding proton and nucleus, $r=\sqrt{B^2+z^2}$.
Following Ref.~\cite{Alvioli:2018jls}, for the ${}^{208} P_b$ nucleus, the density of proton and neutron is described by a Wood-Saxon distribution
\begin{equation}
\label{Wood-Saxon}
\rho^{(p,n)}(r)=\frac{\rho_0^{(p,n)}}{1+e^{(r-R_0^{(p,n)})/a_{(p,n)}}}\,.
\end{equation}
For the neutron density we use $R_0^n=6.7$ fm and $a_n=0.55$ fm \cite{Tarbert:2013jze}. For the proton density we use $R_0^p=6.68$ fm and $a_p=0.447$ fm \cite{Warda:2010qa}. The $\rho_0^{(p,n)}$ parameters are fixed by requiring that the proton and neutron density, integrated over all distance $r$, are normalized  to the number of the protons and neutrons in the lead nucleus, respectively.
\par As already anticipated, the 
DPS1 contribution, the first term in Eq.~(\ref{uno}), stands for the contribution already at work in $pp$ collisions. 
It depends linearly on the nuclear thickness function $T$ and therefore scales as the number of nucleon in the nucleus, $A$. 
\par The second term, the DPS2 contribution, contains in principle 
two-body nuclear distributions. 
We work here in the impulse approximation, neglecting short range correlations  in the nuclei since their contribution may change the results by several percent only~\cite{Alvioli:2019kcy}.
The latter term is therefore proportional to the product of one-body nucleonic densities in the nucleus, \textsl{i.e.} it depends quadratically on $T$ and parametrically scales as $A^{4/3}$.
\par As we already stated above we shall work here for simplicity in the mean field approximation for the nucleon. In such approximation double GPD has a  factorized form :
\begin{equation}
\label{fact}
D^{ij}_p(x_1,x_2,\mu_A,\mu_B,\vec{b}_\perp) \simeq 
f^{i}_p (x_{1},\mu_{A})
f^{j}_p (x_{2},\mu_{B})\,
\mathcal{T}(\vec{b}_\perp)~,
\end{equation}
where the function $\mathcal{T}(\vec{b}_\perp)$ 
describes the probability to find two partons 
at a relative transverse distance $\vec b_\perp$ in the nucleon and  is normalized to unity. In such a simple approximation, this function does not depend on parton flavour and fractional momenta. Then one may define the so-called effective cross section as
\begin{equation}
\sigma_{eff}^{-1} =  \int d \vec{b}_\perp [\mathcal{T}(\vec{b}_\perp)]^2~,
\label{bprofile}
\end{equation}
which controls the double parton interaction rate.
Under all these approximations the DPS cross section in $pA$ collision can be rewritten as
\begin{multline}
\frac{d \sigma^{CD}_{DPS}}{d\Omega_1 d\Omega_2} = \frac{m}{2} 
\sum_{i,j,k,l} 
\sum_{N=p,n} \sigma_{eff}^{-1}
 f^{i}_{p}(x_1) f^{j}_{p}(x_2)
f^{k}_{N}(x_3) f^{l}_{N}(x_4) 
\frac{d \hat \sigma_{ik}^C}{d\Omega_C} \frac{d \hat \sigma_{jl}^D}{d\Omega_D} \int d^2 B \, T_N(B)\,,\\
+ \frac{m}{2} \sum_{i,j,k,l} \sum_{N_3, N_4=p,n}
f^{i}_{p}(x_1) f^{j}_{p}(x_2)
f^{k}_{N_3}(x_3) f^{l}_{N_4}(x_4) 
\frac{d \hat \sigma_{ik}^C}{d\Omega_C} \frac{d \hat \sigma_{jl}^D}{d\Omega_D}\,
\int d^2 B \, T_{N_3}(B) T_{N_4}(B)\,.
\label{due}
\end{multline}
We find important to remark the key observation that leads to the  second term of Eq.~(\ref{uno}): namely that the $b$ and $B$ integrals practically decouple since the nuclear density does not vary on subnuclear scale~\cite{Strikman:2001gz,Calucci:2013pza,BSW}.
As a result this term  depends on $_2$GPDs integrated over transverse 
distance $b_\perp$, \textsl{i.e.} at $\vec \Delta =0$, for which we assume again mean field approximation:
\begin{equation}
\int d \vec{b}_\perp
D^{ij}_{p}(x_1,x_2;\vec{b}_\perp) \simeq f^{i}_p (x_{1}) f^{j}_p (x_{2})\,.\label{61}
\end{equation}
In the DPS1 term, deviations from the mean field approximation for $_2$GPDs are taken into account at least partially by using in our calculations  the experimental value 
of $\sigma_{eff}$ measured in $pp$ collisions. Additional corrections 
of order $10\%-20\%$  to Eq.~(\ref{due})
due to  longitudinal correlations in the nucleon~\cite{BSW} and beyond mean field approximation will be neglected in the following.
Note that after integration in $b_\perp$, this will be the only non-perturbative parameter 
characterising the DPS cross section. We shall neglect small possible dependence of $\sigma_{eff}$ on energy. Indeed while there is some dependence on energy in pQCD and mean field approach, it is at least 
partly compensated by non-perturbative contributions to $\sigma_{eff}$~\cite{BS}.
\par In this last part of the Section we specify the kinematics and additional settings with which we evaluate Eq.~(\ref{due}).
We consider proton lead collisions 
at a centre-of-mass energy
$\sqrt{s_{pN}}$ = 8.16 TeV. Due to the different energies of the proton and lead beams ($E_p = 6.5$ TeV and $E_{Pb} = 2.56$ TeV per nucleon), the resulting proton-nucleon centre-of-mass is boosted with respect to the
laboratory frame by $\Delta y = 1/2 \, \ln E_p/E_N$ = 0.465 in the proton direction, assumed to be at positive rapidity. 
Therefore the muon and jets rapidities, in this frame, 
are given by $y_{CM}=y_{lab}-\Delta y$.
By assuming a rapidity coverage in the laboratory system $|y_{lab}|<2.4$,
this bound translates into the range $-2.865<y_{CM}<1.935$. 
In all calculations, we have always considered 
proton-nucleon centre-of-mass rapidities. 
\par The relevant partonic cross sections have been evaluated 
at leading order~\cite{bookESW} in the respective coupling differential in 
muon and/or jets transverse momenta and rapidities in order to be able to implement realistic kinematical cuts used in experimental analyses. 
For cross sections involving jets, final state partons are identified as jets, as appropriate for a leading
order calculation.  
\par We use~\texttt{CTEQ6L1} free proton parton distributions ~\cite{Pumplin:2002vw} and~\texttt{EPS09} nuclear parton distributions~\cite{Eskola:2009uj}. Consistently with the cross section calculations, both distributions and strong coupling constant have been evaluated at leading order.
We have successfully benchmarked our codes against 
\texttt{DYNNLO}~\cite{Catani:2009sm} and \texttt{ALPGEN}~\cite{Mangano:2002ea}.
The latter has also been used to obtain leading order, parton level, estimates of the SPS  $Wjj$ and $Wb\bar{b}$ cross sections.  
\par  In order to determine the DPS2 contribution we shall use the strategy devised in Ref.~\cite{Alvioli:2019kcy} . The latter exploits the 
experimental capabilities to accurately relate centrality with the impact parameter $B$ of the $pA$ collision.
The procedure for the determination of centrality in $pA$ collisions was  developed  \textsl{e.g.} by ATLAS~\cite{30}. It makes use of the transverse energy $E_T$ 
deposited in the  pseudorapidity interval $-3.2\ge \eta  \ge   -4.9$ (i.e. along the nucleus direction) as a measure of centrality. It was shown in Ref.~\cite{35}     that        $E_T$ in this kinematics is not sensitive to production of hadrons at forward rapidities.  The $E_T$ distribution as a function of  the number of collisions $\nu$ (and thus on the impact parameter $B$) is presented in Refs.~\cite{25, 30, 35} (see also the related discussion in Ref.~\cite{Alvioli:2019kcy}).
\par We close this Section discussing the uncertainties
related to our theoretical predictions. 
As already stated, aside from uncertainties related to  the mean field approximation for  double PDFs, the DPS2 term is largely free of unknowns.
On the contrary, the DPS1 contribution needs, as input, a value for  $\sigma_{eff}$
whose value and the associated error we borrow
from available experimental analyses
and which we propagate to our theoretical predictions.
Theoretical errors due to missing higher orders
can be kept under control by using higher order calculations, which are known for all processes considered in the present paper and are available in the literature. Uncertainties 
related to PDFs and nuclear effects are by far subleading in the present context. Finally, 
when appropriate, we will show statistical 
errors on the predictions  within a given luminosity
scenario, assuming in such a case a Poissonian distribution.
 
\section{Results : $W jj$}
\label{wjj}
\subsection{Kinematics.}
In this Section we present results for the associated production 
of one electroweak boson in one of the scatterings, which then decays leptonically into a muon and a neutrino,
and of a dijet system produced in the other. 
This process has been already analyzed in $pp$ collisions 
at $\sqrt{s}$=7 TeV by ATLAS~\cite{ATLAS_WJJ} and CMS~\cite{CMS_WJJ}
whose results constitute therefore a useful baseline for this analysis. 
For this channel we define the fiducial phase space 
in terms of muon transverse momentum and rapidity
by requiring that $p_{T}^{\mu}>25$ GeV and $|y^{\mu}_{lab}|<2.4$. Additionally the missing transverse energy is required to satisfy $\slashed{E}_T \; > 25$ GeV.
These sets of cuts are the same as those used in the analysis of Ref. \cite{Sirunyan:2019dox}.
The fiducial phase space for jets is defined by  
$p_{T}^{jets}>20$ GeV and $|y^{jets}_{lab}|<2.4$.
As already mentioned, both ATLAS~\cite{ATLAS_WJJ} and CMS~\cite{CMS_WJJ} have measured 
the DPS contribution to the $Wjj$ final state in $pp$ collisions 
at $\sqrt{s}=7$ TeV and found $\sigma_{eff}$ to be:
\begin{eqnarray}
\sigma_{eff}^{ATLAS}  &=& 15 \pm 3 \,(\mbox{stat.}) ^{+5} _{-3} \,(\mbox{syst.}) \,\mbox{mb}\,, \\
\sigma_{eff}^{CMS}  &=& 20.7 \pm 0.8 \, (\mbox{stat.}) \pm 6.6 \,(\mbox{syst.}) \,\mbox{mb}\,. 
\end{eqnarray}
\par We combine these numbers into $\bar{\sigma}_{eff} = 18 \pm 6$  mb. Since we simulate $pA$ collisions
at $\sqrt{s_{pN}}$=8.16 TeV, 
a centre-of-mass energy close to the energies at which those values of $\sigma_{eff}$ have been extracted, we use such an average in our numerical simulations
for the $Wjj$ and $Wb\bar{b}$ final states, neglecting any possible dependence of $\sigma_{eff}$ on energy.
For the DPS contributions, the $W$ cross section has been evaluated with factorization and renormalization scales 
fixed to the $W$ boson mass, $\mu_F=\mu_R=M_W$, while for the dijet cross section to $\mu_F=\mu_R=\sqrt{p_{T,j1}^{2}+p_{T,j2}^{2}}$. The cross section for the SPS $Wjj$ background was 
evaluated with \texttt{ALPGEN}~\cite{Mangano:2002ea} 
with the same selection cuts described above and 
$\mu_F=\mu_R=\sqrt{M_W^2+p_{T,j1}^{2}+p_{T,j2}^{2}}$.
\par 

\subsection{DPS Calculation Results.}
\begin{figure}[t]
\includegraphics[scale=0.60]{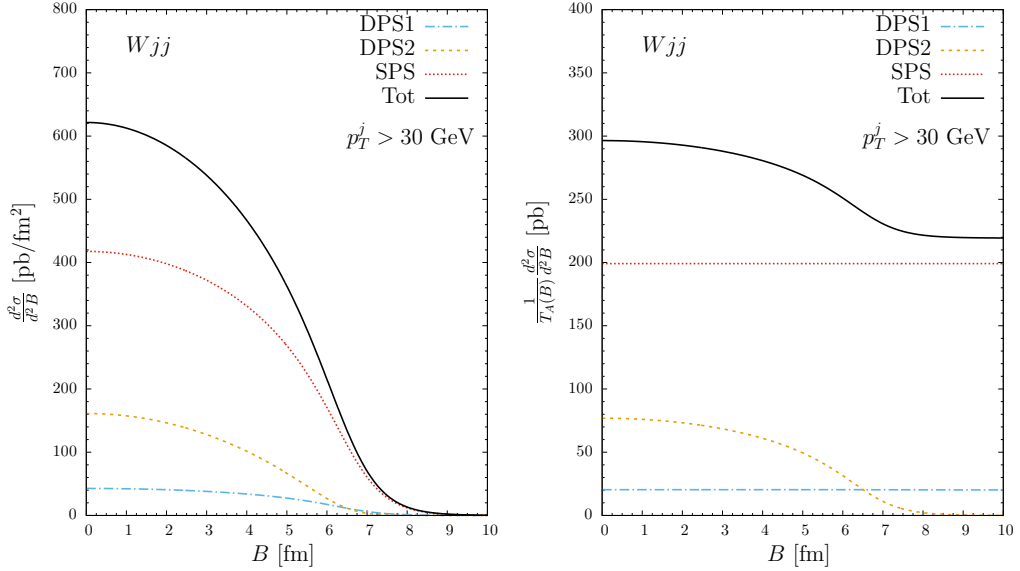}
\caption{\textsl{$Wjj$ DPS cross section as a function of impact parameter $B$ (left panel) and normalized to the nuclear thickness function $T_A(B)$ (right panel).}}
\label{plot:dsdB}
\end{figure}
\setlength{\extrarowheight}{0.2cm}
\begin{table}[t]
\begin{center}
\begin{tabular}{cccc}  \hline  \hline 
 & \hspace{0.5cm} $p_T^j>20$ GeV \hspace{0.5cm} & \hspace{0.5cm} $p_T^j>25$ GeV \hspace{0.5cm} & \hspace{0.5cm} $p_T^j>30$ GeV \\  
$\sigma^{Wjj}$ &  [nb] & [nb] & [nb]\\ \hline
DPS1 & 19 $\pm$ 6  &  8 $\pm$ 3  & 4 $\pm$ 2 \\
DPS2 & 49          &  22       & 11 \\
SPS  & 81          &  57       & 41\\
Tot  & 149 $\pm$ 6  & 87 $\pm$ 3  & 56 $\pm$ 2 \\
\hline
\end{tabular}
\caption{\textsl{Predictions for
$Wjj$ DPS and SPS cross sections in $pA$ collisions in fiducial phase space, for different cuts on jets transverse momenta. 
These numbers refer to charged summed $W$ cross sections accounting for the $W$ boson decaying into muons as well as electrons. The quoted error is entirely due to the propagation of $\bar{\sigma}_{eff}$ uncertainty.}}
\label{Wjj:cs}
\end{center}
\end{table}

\par We are now in position to discuss our results. First we are interested to quantify at the integrated level 
the DPS2 contribution to DPS in $pA$ collisions, which, despite having been predicted theoretically~\cite{Strikman:2001gz}, has not been yet observed experimentally.
\par For this purpose we first report in Tab.~\ref{Wjj:cs} 
the various contributions to the $Wjj$ fiducial cross section. These numbers account for $W$ charged summed cross sections considered in both the muon and electron decay channels. 
From the table it appears that the DPS2 contribution is more than two times larger with respect to DPS1 one.
\par With these numbers at our disposal we may use the strategy put forward in Ref.~\cite{Alvioli:2019kcy} to separate the DPS2 contribution. 
\par We  present in the left panel of Fig.~(\ref{plot:dsdB}) the various contributions to the $Wjj$ cross section differential in impact parameter $B$ for $p_T^j>30$ GeV. 
In the right panel of the same plot the same differential distribution 
is normalized to the nuclear thickness functions.
With such a normalization, the DPS1 contribution will contribute a constant value to the cross section, as well as the SPS background, while DPS2 will show a $B$ dependence driven by $T(B)$. 
The DPS2 observation will essentially rely on the experimental ability to distinguish a non-constant behaviour of such a normalized distribution.  
\par The efficiency of this discrimination method will depend on the accumulated 
integrated luminosity. Here we choose a value in line with data recorded in 2016 $pA$ runs of $\int \mathcal{L} \mbox{dt}= 0.1 \, \mbox{pb}^{-1}$. 
\begin{figure}[t]
\includegraphics[scale=0.60]{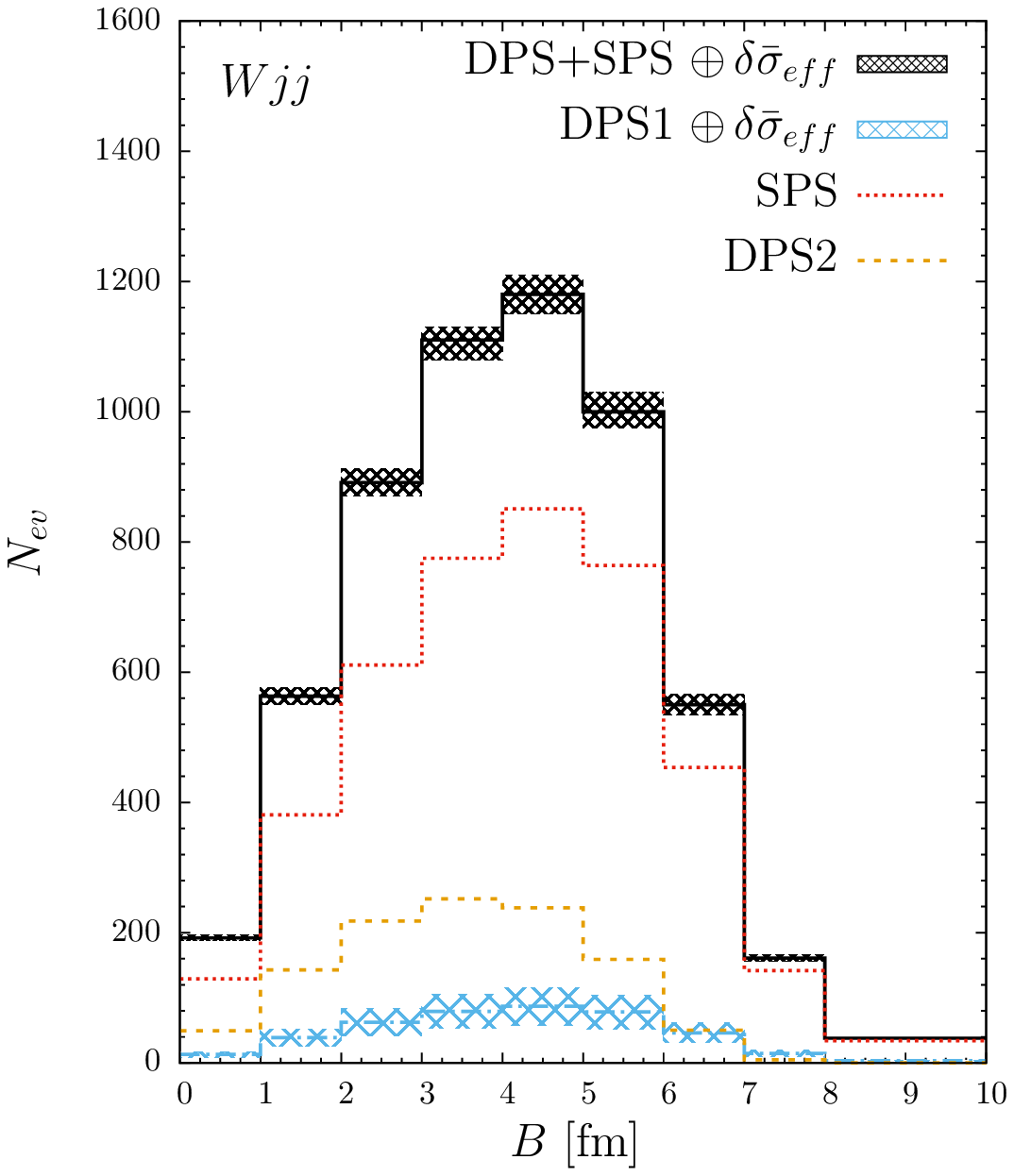}
\includegraphics[scale=0.60]{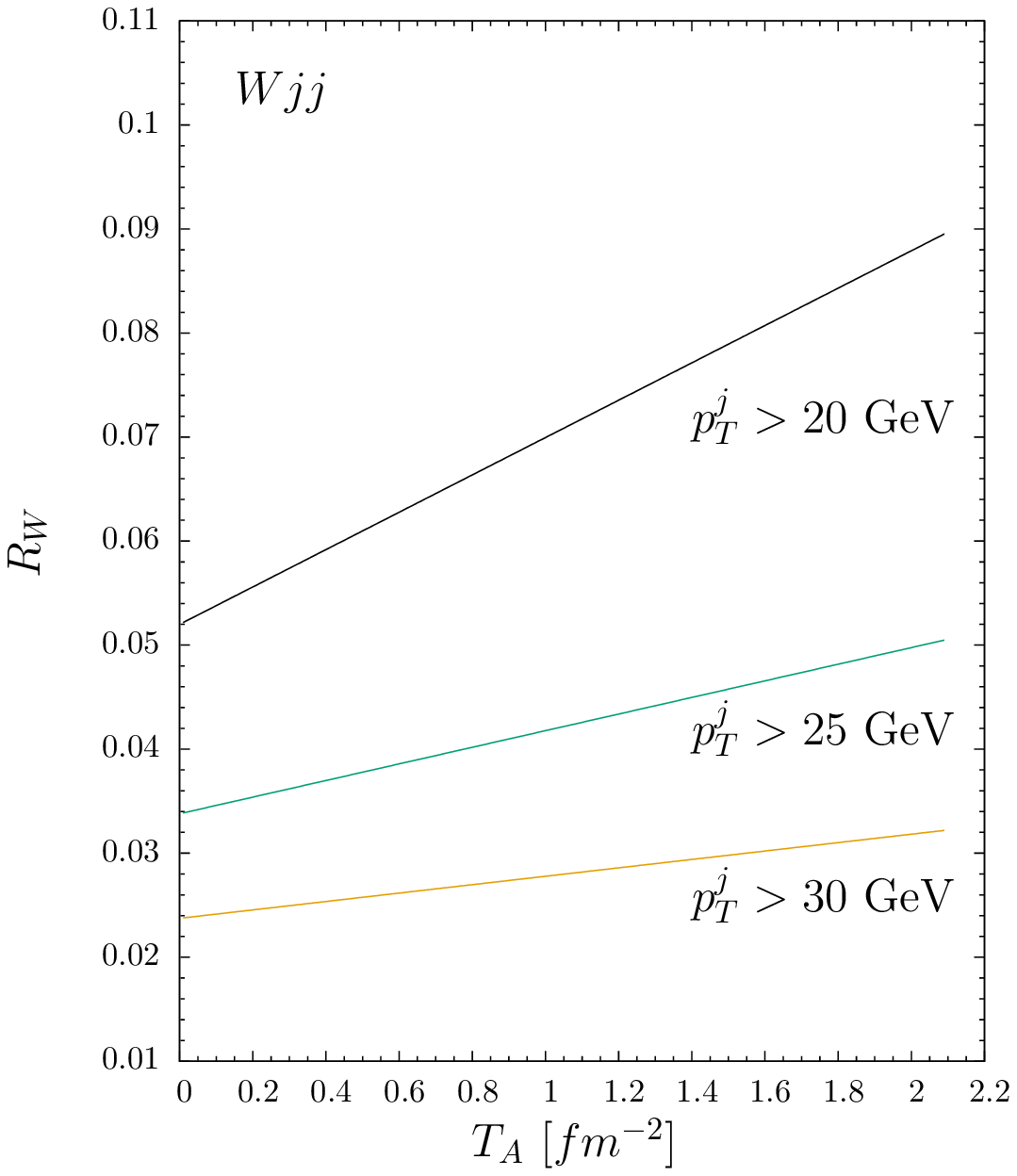}
\caption{\textsl{
Number of events in $Wjj$ final state in bins of $B$ assuming $\int \mathcal{L}dt=0.1 \mbox{pb}^{-1}$ and for $p_T^j>30$ GeV (left). The ratio $R_W$ as a function of $T_A$  for different cuts on jet transverse momenta (right).}}
\label{plot:Nev_wjj}
\end{figure}
\indent For this purpose we present in the left panel of Fig.~(\ref{plot:Nev_wjj})
the number of events for the $Wjj$ final state integrated in bins of $B$. The distribution presents a kinematic zero at $B=0$ due to the jacobian arising from Eq.~(\ref{uno}) when 
the cross section is kept differential in $B$.
On the same plot is also superimposed the uncertainty on the predictions coming from the propagation of the error on $\bar{\sigma}_{eff}$. 
\par The method can be applied to subtract the overwhelming LT SPS contribution, or, at  least to complement the subtraction techniques already developed
in experimental analyses of DPS cross sections. 
For this purpose we consider the ratio  $R_W$ 
between the total number (DPS+SPS) of $Wjj$ events over those for inclusive $W$ production as a function of $T_A(B)$:
\begin{equation}
R_W(T) = N_{Wjj} (T)/ N_W(T).
\label{snow}
\end{equation}
In such a ratio, $N_W(T)$ is linear in $T_A(B)$, as well as the SPS background and DPS1 contribution.
Therefore, in absence of the quadratic DPS2 contribution, 
it would be a constant. Its deviation from such a behaviour will be just due to DPS2 contribution, which will determine the slope of its linear increase.
\par It is worth mentioning that such ratio is directly measurable 
in experiments since $T_A(B)$ is proportional to the number of 
collisions $\nu$  \cite{treleani} and that many of the systematics related to $W$ reconstruction simplify. We present the physical observables as a function  of $T_A(B)$ since they are measured in terms of $T_A(B)$ 
(or equivalently in terms of  a number of collisions  )
in the real
experimental setup  \cite{treleani} .
\par The resulting distribution is presented in the right panel of Fig.~(\ref{plot:Nev_wjj}) for different values of jet transverse momenta cut off. The rise of the slope is related to fast rise of the dijet cross sections entering the DPS2 estimation 
as the cuts on jet transverse momenta are decreased.

\begin{figure}[t]
\includegraphics[scale=0.55]{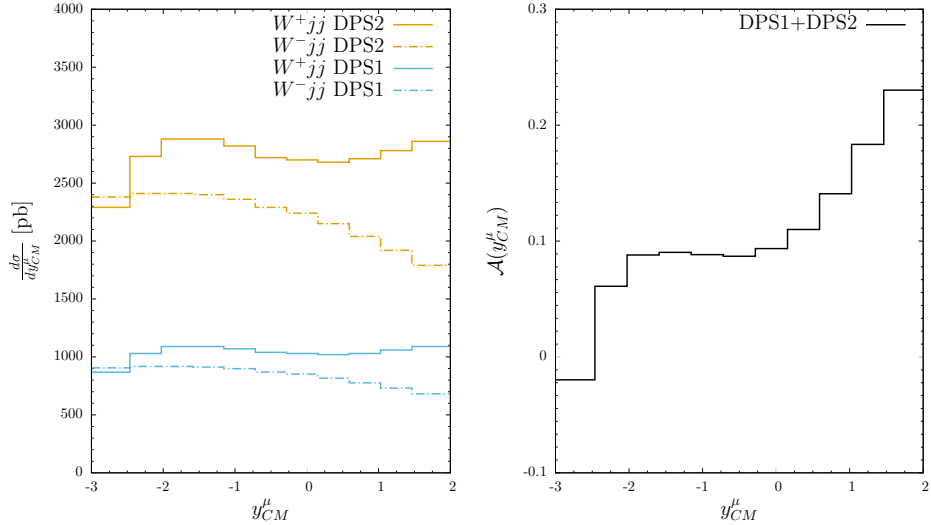}
\caption{\textsl{
$Wjj$ DPS cross sections in $pA$ collisions as a function of the charged muon rapidity in the proton-nucleon centre-of-mass frame (left) and the corresponding muon charge asymmetry (right).}}
\label{plot:dsdeta}
\end{figure}
 \par Given the large number of signal DPS events in the $Wjj$ channel, after proper subtraction of the SPS contribution, 
the characterization of the DPS cross section could be attempted by inspecting the charged lepton 
rapidity distributions. The latter are presented in the left panel of Fig.~(\ref{plot:dsdeta}) for all different charge contribution and DPS mechanism and are obtained 
integrating over impact parameter $B$ and over dijet phase space.
\par Correlations beyond mean field approximation could be appreciated by considering 
the lepton charge asymmetry, a generalization of the familiar observable defined in SPS:
\begin{equation}
\mathcal{A}(y^\mu_{CM})=\frac{d\sigma_{DPS}(W^+jj)-d\sigma_{DPS}(W^-jj)}{d\sigma_{DPS}(W^+jj)+d\sigma_{DPS}(W^-jj)}\,.
\end{equation}
\par The corresponding distribution is presented in the right panel of Fig.~(\ref{plot:dsdeta}).
Given the factorized ansatz for double PDFs and that the dijet system is completely integrated over, 
its line shape is the same as the lepton charge asymmetry measured in SPS production of $W^\pm$ in $pA$ collisions, see for example Fig.~(4) of Ref.~\cite{Sirunyan:2019dox}. 
Therefore, after proper subtraction of SPS and DPS1 contribution, the observation in data of any departure  
from the predicted line shape might be an indication of parton correlations
not accounted for in the mean field approximation.

\subsection{Leading twist background versus DPS2}
\par The main experimental impediment to the observation of the DPS2 contribution is the large LT background. We have calculated the latter in the leading order (LO) approximation by using the \texttt{ALPGEN} generator. 
\begin{figure}[t]
\includegraphics[scale=0.70]{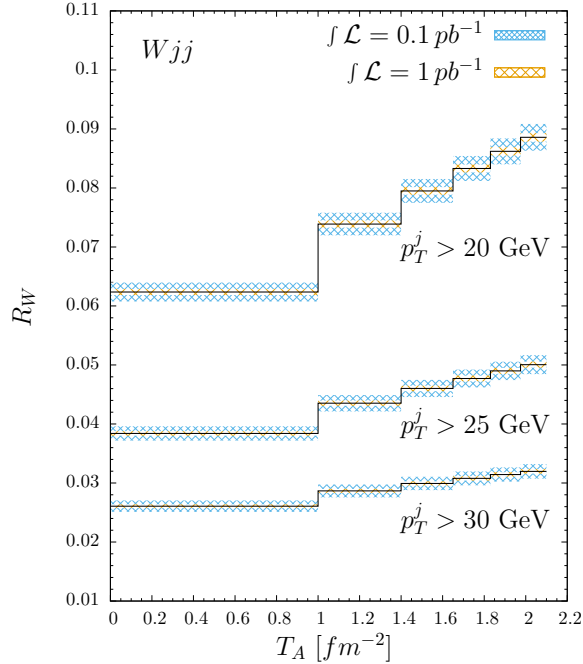}
\caption{\textsl{The ratio $R_W$ as a function of T for $Wjj$ integrated in bins of $T_A$ for two values of the integrated luminosity  and  different cuts on jet transverse momenta.}}
\label{plot:TWjj}
\end{figure}

\par 
We display in the  Fig.~(\ref{plot:TWjj})
the ratio $R_W$,  defined by Eq.~(\ref{snow}) in the previous Subsection, and  integrated in the interval of the thickness function to better facilitate the connection with centrality dependence.
The bin widths are chosen as to evenly distribute
the number of  events across the various $T_A$ bin. 
\par The error band associated to the ratio is obtained assuming a Poissonian 
distribution of statistical errors, obtained from the number 
of total $N_{Wjj}$ and $N_W$ events expected for two values of $\int \mathcal{L} \, dt = 0.1$ and 1 $pb^{-1}$. 
Large (small) values of $T_A$ correspond to central (peripheral) events, with the ratio increasing going from peripheral 
to central collisions. The errors associated to the distribution determine a marked sensitivity to DPS2 mechanism which is responsible for the non zero slope of the distribution. 
\section{Results : $Wb\bar{b}$}
We consider in this Section a special case of the former 
process in which the second scattering produces a 
$b\bar{b}$ heavy-quark pair. 
This particular final state has been analyzed in detail in $pp$ collisions in~\cite{Berger:2011ep} 
where a number of kinematic variables have been proposed to disentangle the signal DPS process from the SPS background. It is worth noticing that this final state is particularly important for new physics searches in $pp$ collisions so that the DPS component needs to be properly modelled. 
For this final state we use $\bar{\sigma}_{eff} = 18 \pm 6$  mb as for the  $Wjj$ case. 
\begin{table}[t]
\begin{center}
\begin{tabular}{cccc}  \hline  \hline 
 & \hspace{0.5cm} $p_T^b>20$ GeV \hspace{0.5cm} & \hspace{0.5cm} $p_T^b>25$ GeV \hspace{0.5cm} & \hspace{0.5cm} $p_T^b>30$ GeV \\ 
$\sigma^{Wb\bar{b}}$ &  [pb] & [pb] & [pb]\\ \hline
DPS1 &  74 $\pm$ 25  & 35 $\pm$ 12 &  18 $\pm$ 6 \\
DPS2 &  196          & 92 &  48 \\
SPS &    234          & 158 &  114 \\
Tot  &   504 $\pm$ 25 & 285 $\pm$ 12 & 180 $\pm$ 6\\
\hline
\end{tabular}
\caption{\textsl{Predictions for
$Wb\bar{b}$ DPS and SPS cross sections in $pA$ collisions in fiducial phase space, for different cuts on jets transverse momenta. These numbers refer to charged summed $W$ cross sections accounting for the $W$ boson decaying into muons as well as electrons. The quoted error is entirely due to the propagation of $\sigma_{eff}$ uncertainty.}}
\label{Wbb:cs}
\end{center}
\end{table}
\begin{figure}[t]
\includegraphics[scale=0.60]{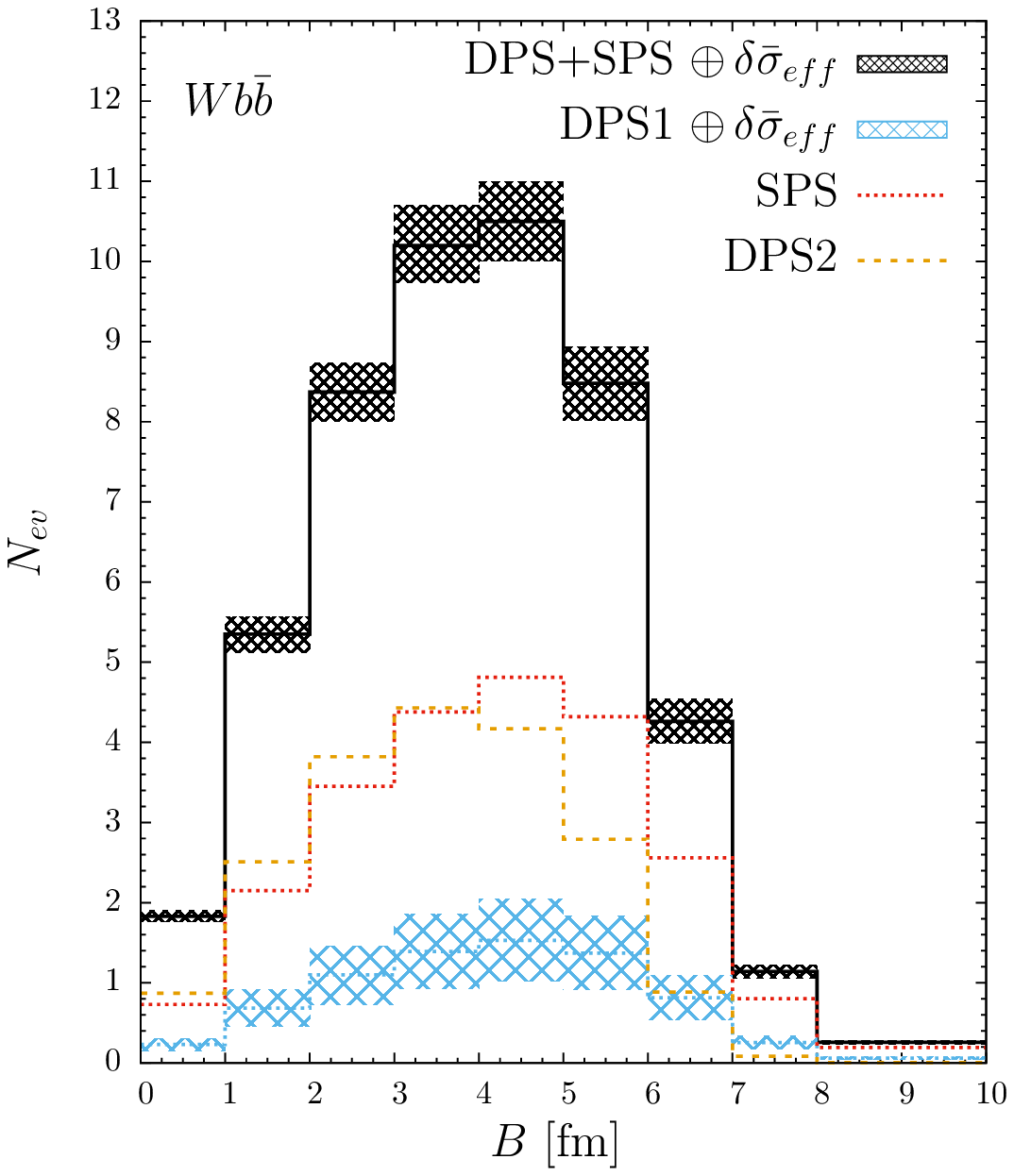}
\includegraphics[scale=0.60]{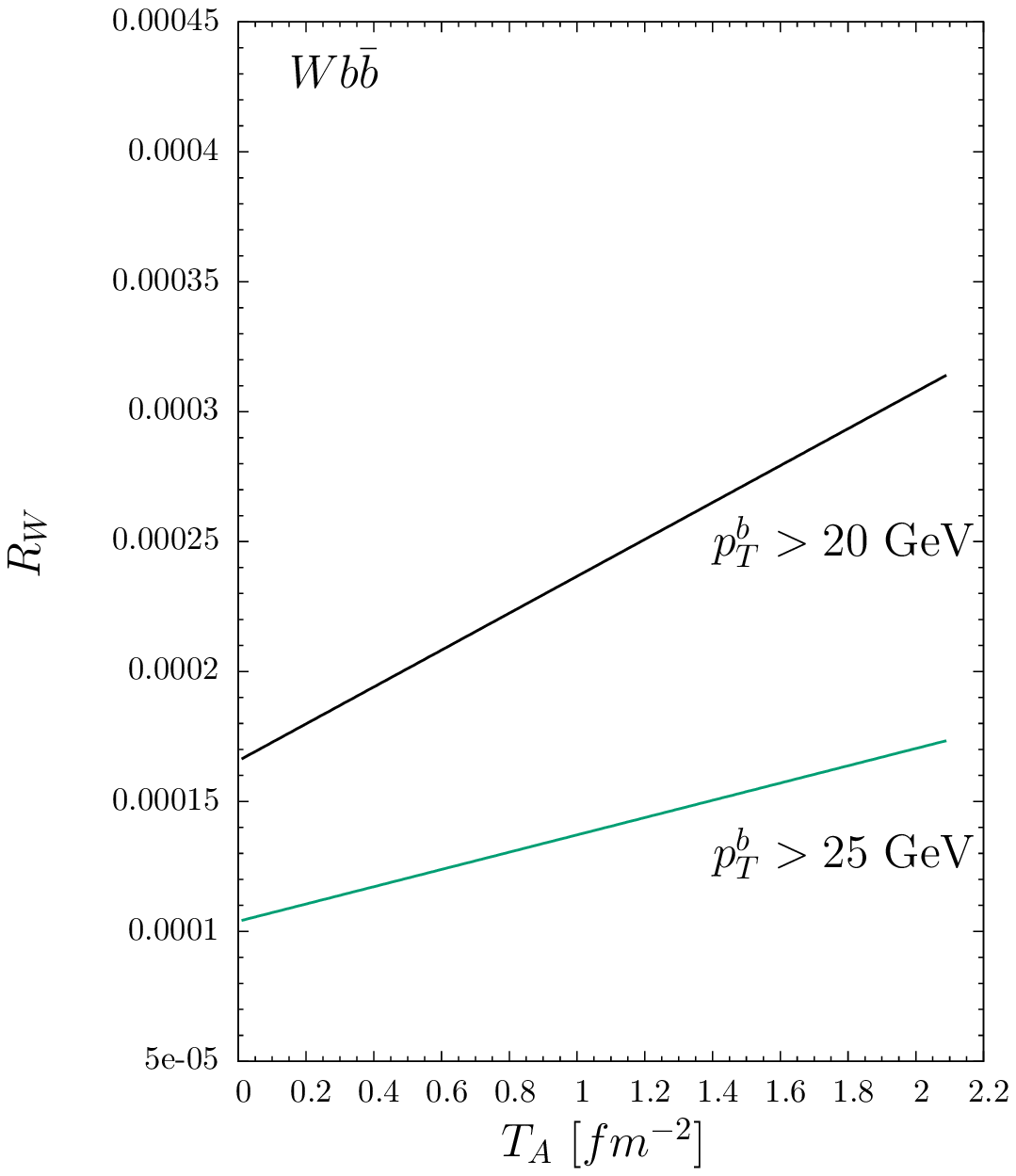}
\caption{\textsl{Number of events in $Wb\bar{b}$ final state in bins of $B$ assuming $\int \mathcal{L}dt=0.1 \mbox{pb}^{-1}$ and for $p_T^j>20$ GeV (left).
The ratio $R_W$ as a function of $T_A$  for different cuts on jet transverse momenta (right).}}
\label{plot:NeV_bb}
\end{figure}

For this channel we define the fiducial phase space 
in terms of muon transverse momentum and rapidity
by requiring that $p_{T}^{\mu}>25$ GeV and $|y^{\mu}_{lab}|<2.4$. Additionally the missing transverse energy is required to satisfy $\slashed{E}_T \; > 25$ GeV.
The fiducial phase space for $b$-jets is given by   
$p_{T}^{b-jets}>20$ GeV and $|y^{b-jets}_{lab}|<2.4$.
For the DPS contributions, the $W$ cross section has been evaluated with factorization and renormalization scales 
fixed to the $W$ boson mass, $\mu_F=\mu_R=M_W$, while for the $b\bar{b}$ dijet cross section to $\mu_F=\mu_R=\sqrt{m_{T,j1}^{2}+m_{T,j2}^{2}}$, 
being $m_T$ the transverse mass of the $b$-jet.
The cross section for the SPS $Wb\bar{b}$ background was 
evaluated with \texttt{ALPGEN}~\cite{Mangano:2002ea} 
with the same selection cuts described above and 
$\mu_F=\mu_R=\sqrt{M_W^2+m_{T,j1}^{2}+m_{T,j2}^{2}}$.

\begin{figure}[ht]
\begin{center}
\includegraphics[scale=0.7]{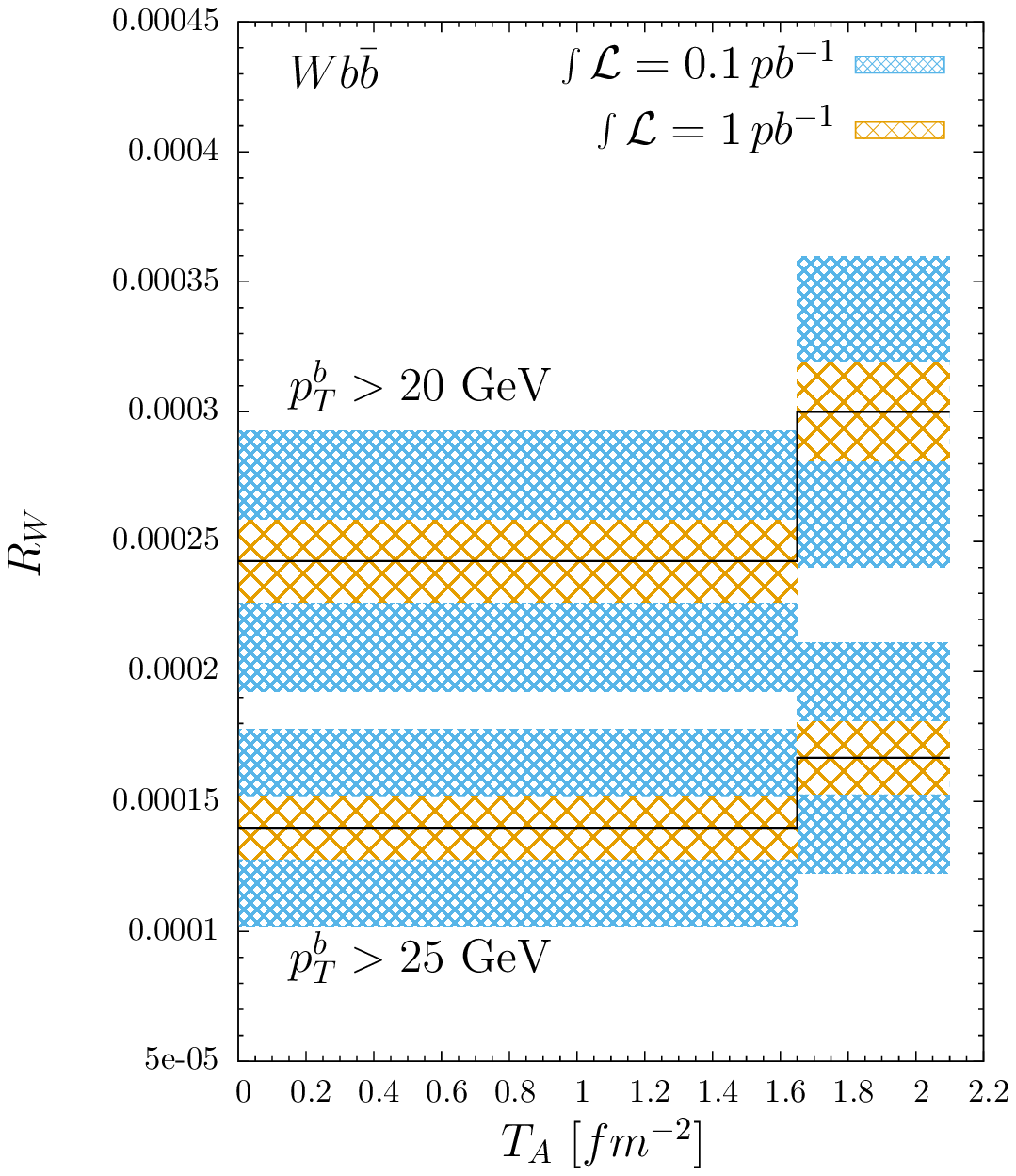}
\caption{\textsl{The ratio $R_W$ as a function of T for $Wb\bar{b}$ final state integrated in bins of $T_A$ for two values of the integrated luminosity and different cuts on jet transverse momenta.}}
\label{plot:Tbin_Wbb}
\end{center}
\end{figure}
The various contributions to the $Wb\bar{b}$ cross sections  are reported in Tab.~\ref{Wbb:cs}.  
As expected, they are reduced by several order of magnitude with respect to the $Wjj$ case. 
Assuming again a rather conservative scenario in which the integrated luminosity is $\int \mathcal{L} \mbox{dt}= 0.1 \, \mbox{pb}^{-1}$, we present in the left panel of Fig.~(\ref{plot:NeV_bb}) the expected 
number of events for the various contributions  
integrated in bins of $B$. 
\par In the right panel of Fig.~(\ref{plot:NeV_bb}) we present, for this particular final state, the ratio $R_W$ between the total number (SPS+DPS) of $Wb\bar b$ events over those for inclusive $W$ production
\begin{equation}
R_W(T) = N_{Wb\bar b} (T)/ N_W(T),
\label{snow1}
\end{equation}
as a function of $T_A(B)$ for different transverse momenta cut off 
on the $b$-jets.

In Fig.~(\ref{plot:Tbin_Wbb}) we present the ratio $R_W$ defined by 
Eq.~(\ref{snow1}) integrated in two ranges
of $T_A$ and assuming $\int \mathcal{L} \mbox{dt}=$ 0.1 and 1 $\mbox{pb}^{-1}$.  The bin widths  were determined from the condition that 
the events are evenly distributed in each bins.

According to our error estimates, the result is not conclusive for the current integrated luminosity scenario, while a sufficient discrimination power is achievable in the  future runs,
although even in the latter case this channel is much less sensitive to DPS2 than the $Wjj$ one.

As already observed in the 
$Wjj$ case, lowering the cut on the transverse momenta of $b$-jet increases the sensitivity to a non constant behaviour of $R_W$.  
\par  
 From the latter plot it is clear that for this final state, given the lower number of events, the identification of a non-constant behaviour in data will be difficult. Nevertheless, since at the $B$-integrated level, the DPS2 contribution is more than twice the DPS1 one, this channel has anyway the potential to allow the  observation of the DPS2 mechanism.

\section{Results : $ssWW$}
\label{Sec:ssWW}
\par Double Drell-Yan like processes 
have been recognized as an ideal laboratory to investigate DPS~\cite{Goebel,mufti}
and its factorization property~\cite{DG_Glauber}.
Among this class of process, the production of a same sign $W$ boson pair ($ssWW$), where each $W$-boson is produced in a distinct hard scattering, has received
special attention~\cite{Kulesza:1999zh,Gaunt:2010pi,
 Golec-Biernat:2014nsa,Ceccopieri:2017oqe,Cotogno:2018mfv,Cao:2017bcb},
since single parton scattering (SPS) at tree-level starts contributing
to higher order in the strong coupling and can be suppressed by additional jet veto requirements.
This process has been investigated in $pA$ collisions in Ref.~\cite{dEnterria:2012jam}.
\par
\begin{figure}[t]
\includegraphics[scale=0.55]{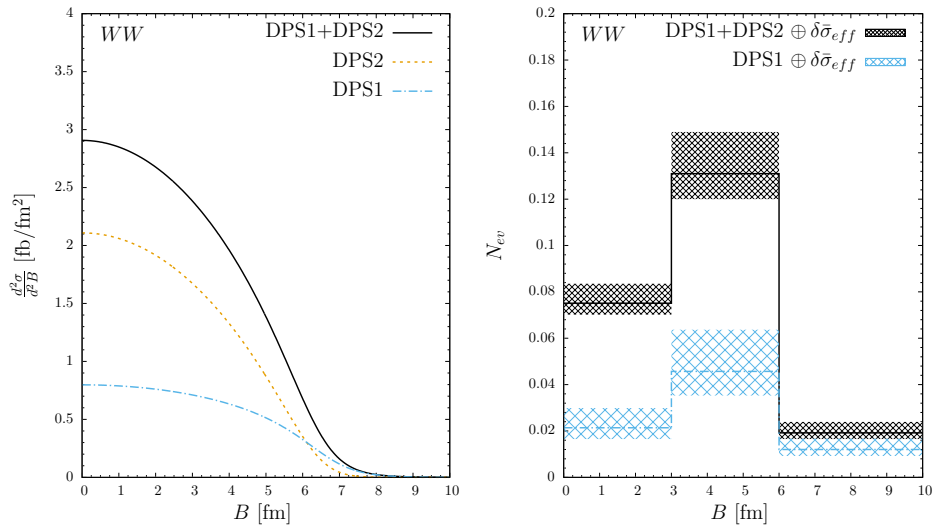}
\caption{\textsl{DPS cross sections as a function of $B$ (left) and expected number of events with $\int \mathcal{L}dt$=1 pb$^{-1}$ in the charged summed dimuon channel (right). The error band represents the propagation of the $\bar{\sigma}_{eff}$ uncertainty.}}
\label{plot:fig6}
\end{figure}

 A measurement of the $ssWW$ DPS cross section in $pp$ collisions at $\sqrt{s}=$13 TeV
has been recently reported  by the CMS collaboration~\cite{Sirunyan:2019zox}. 
In that analysis a value of $\sigma_{eff} = 12.7^{+5.0}_{-2.9}$ mb has been extracted 
and which will be used in our predictions, assuming that such a value is valid also  at $\sqrt{s_{pN}}=$8.16 TeV,
the nominal energy at which we simulate $pA$ collisions in this analysis. 
Again we assumed that its value is the same in both charged 
channels and the same across the fiducial phase space. 
Both $W$'s are required to decay into same sign muons and we adopt 
the fiducial phase space from the analysis of Ref.~\cite{Sirunyan:2019dox}: 
it is given by $p_{T}^{\mu}>25$ GeV for the leading muon,  
$p_{T}^{\mu}>20$ GeV for the subleading one and $|y^{\mu}_{lab}|<2.4$ for muons' rapidities.
Additionally the missing transverse energies are required to satisfy $\slashed{E}_T \; > 25$ and 20 GeV.
\begin{table}[t]
\begin{center}
\begin{tabular}{cccc}  \hline  \hline 
& \hspace{0.5cm} $\sigma^{\mu^+ \mu^+}$ [fb]\hspace{0.5cm} &  \hspace{0.5cm} $\sigma^{\mu^- \mu^-}$ [fb] \hspace{0.5cm} &   $\sigma^{\mu^+ \mu^+}$+  $\sigma^{\mu^- \mu^-}$ [fb]  \\ \hline
DPS1 & \hspace{0.5cm} $48^{+19}_{-11}$  \hspace{0.5cm}  & \hspace{0.5cm} $31^{+12}_{-7}$ \hspace{0.5cm} &  $79^{+31}_{-18}$\\
DPS2 & \hspace{1cm} 88                \hspace{0.5cm}  & \hspace{1cm} 58              \hspace{0.5cm} &  146 \\
DPS  & \hspace{0.5cm} $136^{+19}_{-11}$ \hspace{0.5cm}  & \hspace{0.5cm} $89^{+12}_{-7}$ \hspace{0.5cm} &  $225^{+31}_{-18}$ \\
\hline
\end{tabular}
\caption{Fiducial cross sections for $ssWW$ DPS cross sections for the positive (left), 
negative (central) and charged summed (right) dimuon final state for all DPS contributions.
The quoted errors follow from the propagation of $\sigma_{eff}$ uncertainty.}
\label{ssWW_ref}
\end{center}
\end{table}
We report the cross sections results in Tab.~\ref{ssWW_ref} for various DPS mechanisms 
and for separate dimuon charge configurations. 
In Fig.~\ref{plot:fig6} we present the differential cross sections and the number of expected events 
for $\int \mathcal{L}dt$= 1 pb$^{-1}$, a value within 
reach at future $pA$ runs at LHC. 
\par Considering all leptonic channels ($\mu^\pm \mu^\pm , e^\pm \mu^\pm, e^\pm e^\pm$), the resulting fiducial cross section is four times larger than that reported in 
Tab. \ref{ssWW_ref} and it is of order 1 pb. 
These results are consistent with the ones reported in Ref. \cite{dEnterria:2012jam} after noting
that those have been obtained at higher $\sqrt{s_{pN}}=8.8$ TeV with respect to the one used here 
and that cross sections have been calculated there at next to leading order. 
Given these numbers we conclude that the observation of DPS in this channel will not only depend on the integrated luminosity accumulated in future $pA$ runs but also on the experimental ability to reconstruct $W$'s and its charge via its hadronic decays. 

\section{Conclusions}
\indent In this paper we have calculated 
DPS cross sections for a variety of final states produced in 
$pA$ collisions at the LHC, as well as the corresponding SPS backgrounds. We have discussed a strategy 
to separate the so-called DPS2 contributions, 
pertinent to $pA$ collisions, 
which relies on the experimental capabilities to 
correlate centrality with impact parameter $B$ of 
the proton-nucleus collision.
With this respect 
the $Wjj$ final state has large enough cross sections
to allow the method to be used already with 2016 recorded data. Moreover the distribution 
in lepton charge asymmetry has the potential to uncover correlations in double $_2$GPD of the nucleon beyond the mean field approximation. 
The $Wb\bar{b}$ final state, having lower rate, can still be used at the inclusive level to search for the DPS2 contribution in the future runs, albeit with reduced sensitivity relative to $Wjj$ final state.
The observation of the $ssWW$ final state, being a clean but a rare process,  will depend crucially on the running conditions of the future $pA$ runs and upon the $W$-reconstruction experimental capabilities. 

\begin{acknowledgments}
\noindent
The authors would like to thank  M. Strikman for reading the manuscript and for many useful discussions. The authors would also like to thank Sasha Milov for many useful suggestions and discussions. 
The work was supported by Israel Science  Foundation  under  the  grant  2025311.
The diagram in this paper has been drawn with Jaxodraw package version 2.0 \cite{Binosi:2008ig}.
\end{acknowledgments}


\begin{thebibliography}{99}
\bibitem{TreleaniPaver82}
N.\ Paver and D.\ Treleani,
  Nuovo Cim.\  A {\bf 70} (1982) 215.

\bibitem{mufti} M.\ Mekhfi, Phys. Rev. D{\bf 32}, 2371 (1985).

\bibitem{stirling} J.R.\ Gaunt and W.J.\ Stirling,
	  JHEP {\bf 1003}, 005 (2010)   
	 
\bibitem{BDFS1}
  B.\ Blok, Yu.\ Dokshitzer, L.\ Frankfurt and M.\ Strikman,
  Phys.\ Rev.\  D {\bf 83}, 071501 (2011)
  
 \bibitem{Diehl} M.~Diehl,
	  PoS D {\bf IS2010} (2010) 223

\bibitem{stirling1} J.R.\ Gaunt and W.J.\ Stirling,
	  JHEP {\bf 1106},  048 (2011) 
	  
  \bibitem{BDFS2} B.\ Blok, Yu.\ Dokshitser, L.\ Frankfurt and M.\ Strikman,
  Eur.\ Phys.\ J.\ C {\bf72}, 1963  (2012)
  
\bibitem{Diehl2} M.\ Diehl, D.\ Ostermeier and A.\ Schafer,
	  JHEP {\bf 1203} (2012) 089

\bibitem{BDFS3} B.\ Blok, Yu.\ Dokshitser, L.\ Frankfurt and M.\ Strikman,
 arXiv:1206.5594v1 [hep-ph] (unpublished).
 \bibitem{BDFS4}
 B.~Blok, Y.~Dokshitzer, L.~Frankfurt and M.~Strikman,
  Eur.\ Phys.\ J.\ C {\bf 74} (2014) 2926


\bibitem{Diehl:2017kgu}
  M.~Diehl, J.~R.~Gaunt and K.~Schönwald,
  JHEP {\bf 1706} (2017) 083

\bibitem{Manohar:2012jr}

  A.~V.~Manohar and W.~J.~Waalewijn,
  Phys.\ Rev.\ D {\bf 85} (2012) 114009
  
\bibitem{BSW}
  B.~Blok, M.~Strikman and U.~A.~Wiedemann,
  Eur.\ Phys.\ J.\ C {\bf 73} (2013) no.6,  2433

\bibitem{book} Adv.\ Ser.\ Direct.\ High Energy Phys.\  {\bf 29} (2018) 2019, P. Bartalini and J. Gaunt Editors.

  \bibitem{ATLAS_WJJ} 
  G.~Aad {\it et al.} [ATLAS Collaboration],
  New J.\ Phys.\  {\bf 15} (2013) 033038
  
 \bibitem{CMS_WJJ} 
  S.~Chatrchyan {\it et al.} [CMS Collaboration],
  JHEP {\bf 1403} (2014) 032
  
\bibitem{Belyaev} I.~M.~Belyaev, talk at MPI-2015 conference.

\bibitem{LHCb}R.~Aaij  {\it et al.} [LHCb Collaboration],  JHEP, 1206 (2012) 141; Addendum 1403 (2014) 108 
 
\bibitem{LHCb1}R.~Aaij  {\it et al.} [LHCb Collaboration], Nucl. Phys. B871 (2013) 1.

\bibitem{LHCb2} R.~Aaij {\it et al.} [LHCb Collaboration],
  JHEP {\bf 1607} (2016) 052

  
\bibitem{Sirunyan:2019zox}
  A.~M.~Sirunyan {\it et al.} [CMS Collaboration],
  Eur.\ Phys.\ J.\ C {\bf 80} (2020) no.1,  41
  
\bibitem{pythia}T.~Sjöstrand {\it et al.},
  Comput.\ Phys.\ Commun.\  {\bf 191} (2015) 159
  %
\bibitem{herwig}
  M.~Bahr {\it et al.},
  Eur.\ Phys.\ J.\ C {\bf 58} (2008) 639
\bibitem{calucci} 
  G.~Calucci and D.~Treleani,
  \textsl{Phys.~Rev.~}{\bf D60}, 054023 (1999).

\bibitem{Rinaldi:2018slz}
  M.~Rinaldi and F.~A.~Ceccopieri,
  Phys.\ Rev.\ D {\bf 97} (2018) no.7,  071501
 
  \bibitem{Strikman:2001gz}
  M.~Strikman and D.~Treleani,
  Phys.\ Rev.\ Lett.\  {\bf 88} (2002) 031801

\bibitem{Helenius:2019uge}
  I.~Helenius and H.~Paukkunen,
  Phys.\ Lett.\ B {\bf 800} (2020) 135084

 \bibitem{sde}D.~d'Enterria and A.~Snigirev,
  Adv.\ Ser.\ Direct.\ High Energy Phys.\  {\bf 29} (2018) 159

\bibitem{Alvioli:2019kcy}
  M.~Alvioli, M.~Azarkin, B.~Blok and M.~Strikman,
  Eur.\ Phys.\ J.\ C {\bf 79} (2019) no.6,  482
  
\bibitem{Berger:2011ep}
  E.~L.~Berger {\it et al.}, 
  Phys.\ Rev.\ D {\bf 84} (2011) 074021
\bibitem{Kulesza:1999zh}
  A.~Kulesza and W.~J.~Stirling,
  Phys.\ Lett.\ B {\bf 475}, 168 (2000).
  
\bibitem{Maina:2009sj}
  E.~Maina,
  JHEP {\bf 09}, 081 (2009).

\bibitem{Gaunt:2010pi}
  J.~R.~Gaunt \textit{et al.}, 
  Eur.\ Phys.\ J.\ C {\bf 69}, 53 (2010).
  
\bibitem{Alvioli:2018jls}
  M.~Alvioli and M.~Strikman,
  Phys.\ Rev.\ C {\bf 100} (2019) no.2,  024912
  
\bibitem{Tarbert:2013jze}
  C.~M.~Tarbert {\it et al.},
  Phys.\ Rev.\ Lett.\  {\bf 112} (2014) no.24,  242502

\bibitem{Warda:2010qa}
  M.~Warda, X.~Vinas, X.~Roca-Maza and M.~Centelles,
  Phys.\ Rev.\ C {\bf 81} (2010) 054309
  
\bibitem{Calucci:2013pza}
  S.~Salvini, D.~Treleani and G.~Calucci,
  Phys.\ Rev.\ D {\bf 89} (2014) no.1,  016020


  \bibitem{BS} B.~Blok and M.~Strikman,
  Phys.\ Lett.\ B {\bf 772} (2017) 219

\bibitem{bookESW} 
R.~K.~Ellis, W.~J.~Stirling,  B.~R.~Webber,
QCD and Collider Physics, Cambridge University Press

\bibitem{Pumplin:2002vw}
  J.~Pumplin \textit{et al.}, 
  JHEP {\bf 0207} (2002) 012


\bibitem{Catani:2009sm}
  S.~Catani, L.~Cieri, G.~Ferrera, D.~de Florian and M.~Grazzini,
  Phys.\ Rev.\ Lett.\  {\bf 103} (2009) 082001
  
\bibitem{Mangano:2002ea}
  M.~L.~Mangano, M.~Moretti, F.~Piccinini, R.~Pittau and A.~D.~Polosa,
  JHEP {\bf 0307} (2003) 001



\bibitem{Eskola:2009uj}
  K.~J.~Eskola, H.~Paukkunen and C.~A.~Salgado,
  JHEP {\bf 0904} (2009) 065


\bibitem{Sirunyan:2019dox}
  A.~M.~Sirunyan {\it et al.} [CMS Collaboration],
  Phys.\ Lett.\ B {\bf 800} (2020) 135048
  
\bibitem{25}
  G.~Aad {\it et al.} [ATLAS Collaboration],
  Phys.\ Lett.\ B {\bf 763} (2016) 313
 
 
 
\bibitem{30}
  G.~Aad {\it et al.} [ATLAS Collaboration],
  Eur.\ Phys.\ J.\ C {\bf 76} (2016) no.4,  199




\bibitem{35}
  G.~Aad {\it et al.} [ATLAS Collaboration],
  Phys.\ Lett.\ B {\bf 756} (2016) 10
 \bibitem{treleani} M.~L.~Miller, K.~Reygers, S.~J.~Sanders and P.~Steinberg,
  Ann.\ Rev.\ Nucl.\ Part.\ Sci.\  {\bf 57} (2007) 205
  doi:10.1146/annurev.nucl.57.090506.123020
   
  
\bibitem{Goebel}
  C.~Goebel, F.~Halzen and D.~M.~Scott,
  Phys.\ Rev.\ D {\bf 22}, 2789 (1980).

    \bibitem{DG_Glauber}
  M.~Diehl \textit{et al.}
  JHEP {\bf 1601}, 076 (2016).




   
\bibitem{Cao:2017bcb}
  Q.~H.~Cao, Y.~Liu, K.~P.~Xie and B.~Yan,
  Phys.\ Rev.\ D {\bf 97} (2018) no.3,  035013
  
  
\bibitem{Cotogno:2018mfv}
  S.~Cotogno, T.~Kasemets and M.~Myska,
  Phys.\ Rev.\ D {\bf 100} (2019) no.1,  011503
  
 

  
\bibitem{Ceccopieri:2017oqe}
  F.~A.~Ceccopieri, M.~Rinaldi and S.~Scopetta,
  Phys.\ Rev.\ D {\bf 95} (2017) no.11,  114030
 


  
\bibitem{Golec-Biernat:2014nsa}
  K.~Golec-Biernat and E.~Lewandowska,
  Phys.\ Rev.\ D {\bf 90} (2014) no.9,  094032
  

\bibitem{dEnterria:2012jam}
  D.~d'Enterria and A.~M.~Snigirev,
  Phys.\ Lett.\ B {\bf 718} (2013) 1395




\bibitem{Binosi:2008ig}
  D.~Binosi, J.~Collins, C.~Kaufhold and L.~Theussl,
  Comput.\ Phys.\ Commun.\  {\bf 180} (2009) 1709
  
  
  \end{thebibliography}
  \end{document}